\documentstyle{elsart}
\input epsf
\begin{document}
\begin{frontmatter}
\title{Ensemble properties of securities traded in the NASDAQ market}

\author{Fabrizio Lillo\thanksref{mail1}}
and
\author{Rosario N. Mantegna}

\address{Istituto Nazionale per la Fisica della Materia, Unit\`a di Palermo\\
and\\ Dipartimento di Fisica e Tecnologie Relative, Universit\`a 
di Palermo, Viale delle Scienze, I-90128, Palermo, Italia}

\thanks[mail1]{corresponding author, lillo@lagash.dft.unipa.it}

\begin{abstract}
We study the price dynamics of stocks traded in the NASDAQ
market by considering the statistical properties of an ensemble 
of stocks traded simultaneously. 
For each trading day of our database, we study the
ensemble return distribution by extracting its first two central moments.
According to previous results obtained for the NYSE market, we find that
the second moment is a long-range correlated variable. We compare 
time-averaged and ensemble-averaged price returns and we show that the 
two averaging procedures lead to different statistical results. 

\end{abstract}


\end{frontmatter}

\section{Introduction}
In recent years physicists started to interact with economists to 
concur to the modeling of financial markets as
model complex systems \cite{Ms99,Bp00}. 
A reliable model of a financial market should be able to reproduce and
to explain the stylized facts observed in the real markets.  
These stylized facts mainly   
refer to the statistical properties of asset returns
and volatility and to the degree and nature of 
cross-correlation between different assets traded 
synchronously or quasi synchronously and belonging to
given portfolios. Recently, we have proposed to
look at a different aspect of an ensemble of stock by considering 
the statistical properties (shape, moments, etc.) of the 
ensemble return distribution of stocks simultaneously traded
in a market \cite{Lillo99,epjb00,variety00,liegi}. Our studies  
have shown that some of the statistical properties
of the ensemble return distribution of stocks traded in the NYSE market 
are not described by simple market models as the single-index model 
\cite{Elton,Campbell}. A similar conclusion has been reached in Ref.
\cite{jpb}.
In this paper we study the statistical properties of the ensemble
return distribution of the NASDAQ market in the 12-year period 1987-1998. 
The main motivation for this study is to compare our previous findings 
obtained for the NYSE with the empirical results observed in a different market.
The NASDAQ market is very different from the NYSE. Stocks traded in the NASDAQ
are usually more volatile (mean volatility is $4.6 \%$ 
per year) than those traded in the NYSE (mean volatility is $2.4 \%$ per year).    
In 1987 the NASDAQ was a relatively small market but its total capitalization
and the number of traded companies increased very fast and in 1994 NASDAQ 
surpassed the NYSE in annual share volume.

\section{Ensemble return distribution properties for the NASDAQ}

The investigated market is the NASDAQ 
during the 12-year period from January 1987 to December 1998 
which corresponds to 3032 trading days.  
We consider the ensemble of all stocks traded in the NASDAQ. 
The number of stocks traded in the NASDAQ is increasing
in the investigated period and it ranges from $864$ at 
the beginning of 1987 to $4280$ at the end of 1998.
The total number of data records exceeds $6$ millions. 
The variable investigated in our analysis is the daily price 
return, which is defined as
\begin{equation}
R_i(t)\equiv\frac{Y_i(t+1)-Y_i(t)}{Y_i(t)},
\end{equation}  
where $Y_i(t)$ is the closure price of $i-$th stock at 
day $t$ ($t=1,2,..$). For each trading day $t$, 
we consider $n_t$ returns, where $n_t$ is the 
total number of stocks traded in the NASDAQ at the selected
day $t$. In our study we use a ``market time". With this choice, 
we consider only the trading days and we remove the weekends 
and the holidays from the calendar time.
A database of more than 6 millions records unavoidably 
contains some errors. A direct control of a so large 
database is not realistic. For this reason, to avoid 
spurious results we filter the data by not considering 
daily price returns which are in absolute values greater than 
$50\%$. We extract the $n_t$ 
returns of the $n_t$ stocks for each trading day $t$. The 
probability density function (pdf) of these
returns $P_t(R)$ provides  information about the kind of activity
occurring in the market at the selected trading day $t$.    
Figure 1 shows the contour plot of the logarithm of the pdf 
as a function of the return and of the trading day.
In Fig. 1 there are long time periods, see for example the 
three-year period 1993-1995, in which the 
central part of the distribution maintains its shape and the 
equiprobability contour lines are approximately parallel one 
to each other. On the other hand there are 
time periods in which the shape of the distribution changes 
drastically. In general these periods corresponds 
to financial turmoil in the market \cite{epjb00}.
\begin{figure}[tc]
\epsfxsize=4.1in
\centerline{\epsfbox{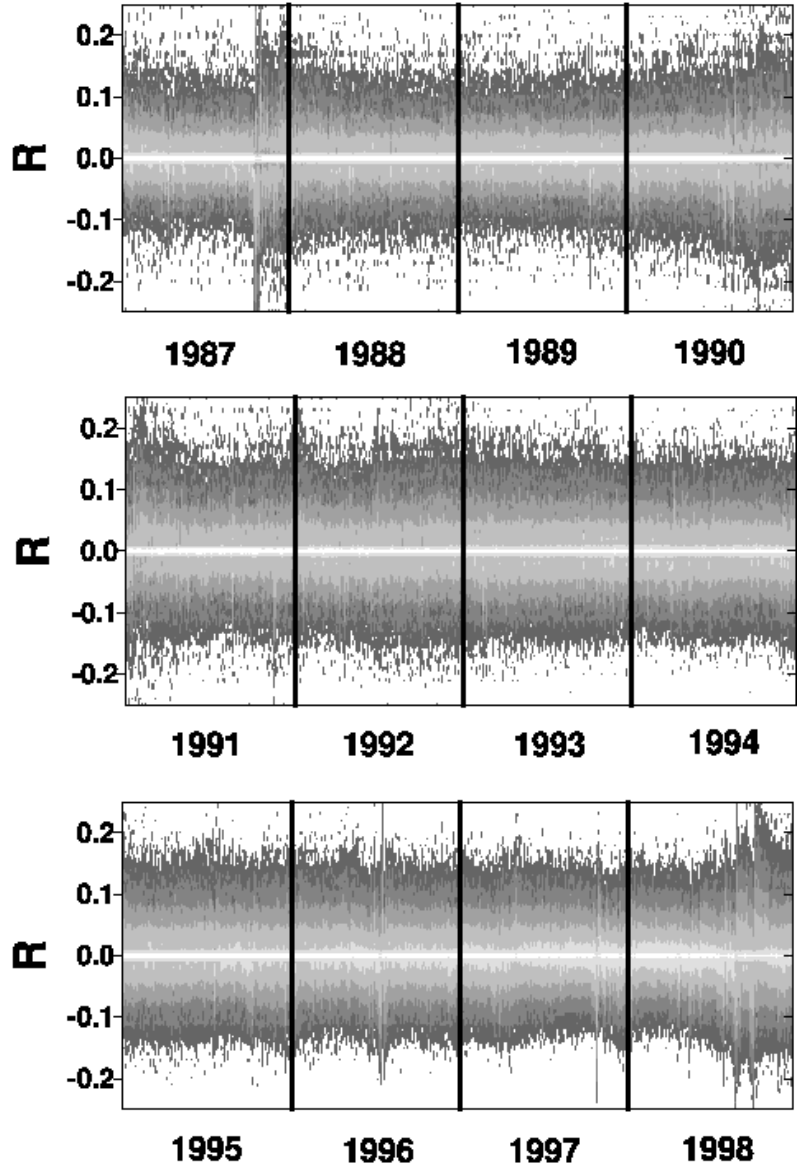}}
\vspace{0.5cm}
\caption{Contour plot of the logarithm of the ensemble return distribution
for the 12-year investigated period from January 1987 to December 1998. 
The contour plot is obtained for equidistant 
intervals of the logarithmic probability density. The brightest area of 
the contour plot corresponds to the most probable value.}
\label{fig1}
 \end{figure}
In order to characterize more quantitatively  the ensemble return
distribution at day $t$, we extract the  first two central moments
at each of the $3032$  trading days. Specifically, we consider the
mean and the standard deviation of $R_i(t)$ defined as
\begin{eqnarray}
&&\mu (t)=\frac{1}{n_t}\sum_{i=1}^{n_t} R_i(t), \\
&&\sigma (t)= \sqrt{\frac{1}{n_t}\left(\sum_{i=1}^{n_t}
(R_i(t)-\mu(t))^2\right)}.
\end{eqnarray}
The mean of price returns $\mu(t)$ quantifies the general trend
of the market at day $t$.  The standard deviation $\sigma(t)$
gives a measure of the width of  the ensemble return distribution.
We call this quantity {\it variety} \cite{Lillo99,variety00}
of the  ensemble because it
gives a measure of the variety of behavior observed in a
financial market at a given day. A large value of $\sigma(t)$
indicates that different companies are characterized by rather
different returns at day $t$.  The mean and the standard
deviation of price returns are not constant and fluctuate in time.
The probability density function of the mean $\mu(t)$ is leptokurtic
because of the correlation between stocks.   
In agreement with previous results on the NYSE market \cite{Lillo99,variety00}, 
we find that the mean $\mu(t)$ is a random variable with very short 
time memory, 
whereas the autocorrelation function of $\sigma(t)$ is a slow decaying function
and lacks a typical time scale.
\begin{figure}[tc]
\epsfxsize=4.0in
\centerline{\epsfbox{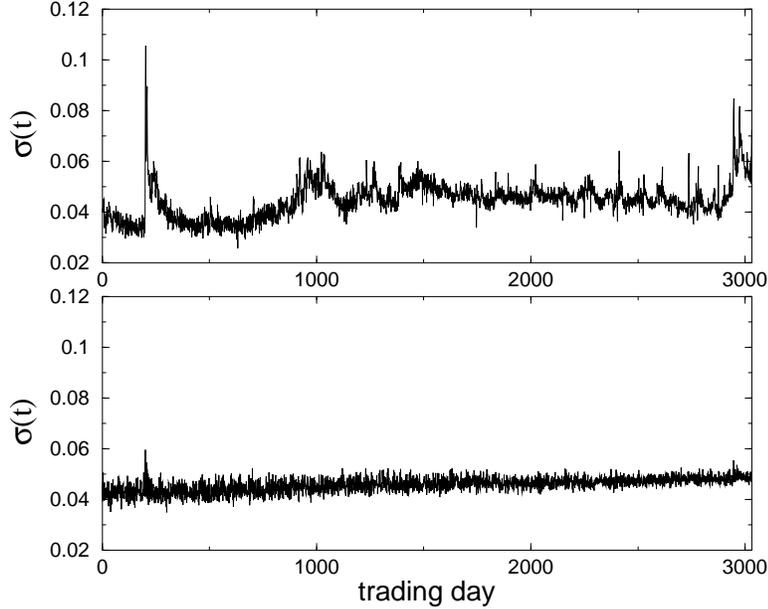}}
\vspace{0.3cm}
\caption{Time evolution of the variety $\sigma(t)$ for the NASDAQ 
market (top) and for a surrogate market generated according to the non 
Gaussian single index model of Eq. (6) (bottom).}
\label{fig2}
\end{figure}
Figure 2 shows the time series of the variety $\sigma(t)$ for the NASDAQ.
The time series of the variety is non stationary and shows several bursts 
of activity and relatively long time periods in which the variety 
has a slow dynamics.
The effects of these observations are reflected in the properties of the
autocorrelation function.
We observe that the autocorrelation function of the variety is greater
than $0.25$ after $100$ trading days.
\begin{figure}[tc]
\epsfxsize=5.0in
\centerline{\epsfbox{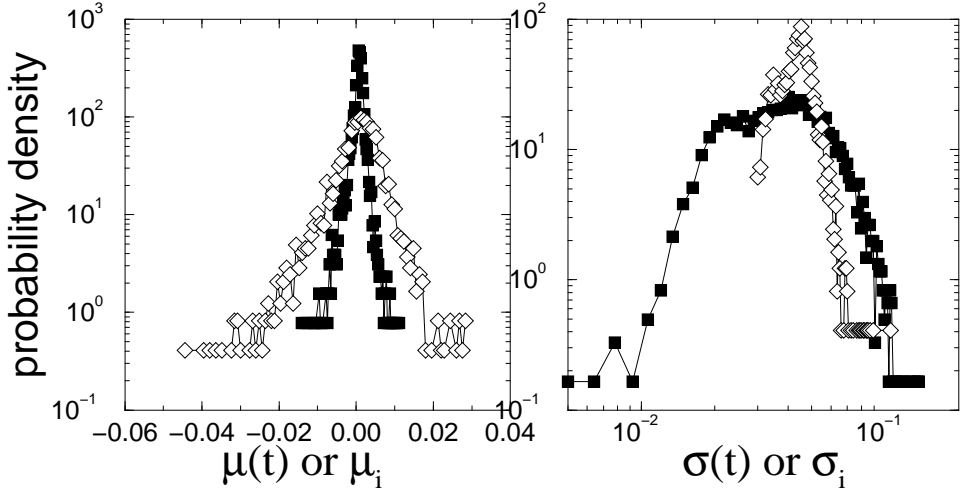}}
\vspace{0.5cm}
\caption{In the left panel we show the probability density function of the
mean $\mu(t)$ of the ensemble return distribution (white diamond) 
and of the mean of the daily return $\mu_i$ of all the stocks traded in the 
NASDAQ (black square). (right) The right panel is the probability density 
function of the variety $\sigma(t)$ , i.e. the  variance of the 
ensemble return distribution (white diamond) and of the volatility
$\sigma_i$, i.e. the variance of the daily return, of the all the
stocks  traded in the NASDAQ (black square).}
\label{fig3}
 \end{figure}

It has been shown by us \cite{variety00} that information about the relative
strength of cross-correlation between different stocks and the autocorrelation
of price returns can be obtained by comparing the statistical 
properties of time-averaged and ensemble-averaged
quantities. To this end for each stock traded in the NASDAQ we extract the 
first two central moments of the time series $R_i(t)$ defined as
\begin{eqnarray}
&&\mu_i=\frac{1}{t^{b}_i-t^{a}_i}\sum_{t=t^{a}_i}^{t^{b}_i} R_i(t), \\
&&\sigma_i= \sqrt{\frac{1}{t^{b}_i-t^{a}_i}
\left(\sum_{t=t^{a}_i}^{t^{b}_i} (R_i(t)-\mu_i)^2\right)},
\end{eqnarray} 
where $t^{a}_i$ indicates the first trading day of our database or 
the day at which the asset $i$ entered the market and $t^{b}_i$ 
indicates the last trading day of our database or 
the day at which the asset $i$ quit the NASDAQ market.
The quantity $\mu_i$ gives 
a measure of the overall performance of stock $i$ 
in the period. The standard deviation $\sigma_i$ is 
called {\it historical volatility} in the financial literature
and quantifies the risk associated with the $i$-th stock. 
This quantity is of primary importance in risk management 
and in option pricing. 
The left panel of Figure 3 shows the
pdf of the time-averaged mean $\mu_i$ and the pdf of the 
ensemble-averaged mean $\mu(t)$.  The pdf of $\mu_i$ is
non-Gaussian and it is much  more peaked than the pdf of $\mu(t)$.
Hence the statistical behavior observed by investigating a large
ensemble in a market day is not representative of the statistical
behavior observed by investigating the time evolution of single
stocks. 
This comparison can be performed also for the second moment of the
distributions. In the right panel of Figure 3
we compare the pdf of the volatility
$\sigma_i$ and the pdf of the variety $\sigma(t)$. Also in this
case, the statistical properties of $\sigma_i$  and $\sigma(t)$ are
different. Specifically, the pdf of $\sigma(t)$  is more peaked
than the pdf of $\sigma_i$.
We showed \cite{variety00}  that the width of the pdf of $\mu(t)$ 
is related to the mean
synchronous cross-covariance between pairs of stock returns, whereas
the width of the pdf of $\mu_i$ is related to the mean  
autocorrelation of stock return time series. The left panel of 
Figure 3 confirms that
also for the NASDAQ market the synchronous cross-correlations between 
the stocks are on average 
stronger than the single stock correlation present in the 
whole portfolio at two different trading day.  
In order to test how robust are these conclusion on the time horizon
used to define returns, we perform the same analysis by considering 
weekly (5 trading days) returns and we find the same discrepancy between
the pdfs of time averaged and ensemble averaged quantities.
In the following we compare our results with the results obtained by modeling
the price dynamics with a
single index model \cite{Elton,Campbell}. The single-index 
model assumes that the returns of all assets are controlled by one 
factor. 
For any asset $i$, we have
\begin{equation}
R_i(t)=\alpha_i+\beta_i R_M(t)+\epsilon_i(t),
\end{equation}
where $R_i(t)$ and $R_M(t)$ are the return of the asset $i$ and of the 
market factor at day $t$, respectively, $\alpha_i$ and $\beta_i$ are two 
real parameters and $\epsilon_i(t)$ is a zero mean noise term 
characterized by a variance equal to $\sigma^2_{\epsilon_i}$.
Our choice for the market factor is the NASDAQ 100 index and we assume 
that $\epsilon_i=\sigma_{\epsilon_i}w$, where $w$ is a random variable
distributed according to a Student's {\it t} density function with
exponent such that $P(w)\sim w^{-4}$.
The use of a non Gaussian noise term has been recently proposed in 
Ref. \cite{jpb}. 
We estimate the model parameters for each asset and we generate an 
artificial market according to Eq. (6). A similar analysis has been 
performed in Ref. \cite{variety00} for the NYSE market. 
We find that the single index model explains well the statistical 
properties of $\mu_i$, $\mu(t)$ and $\sigma_i$ but fails in describing 
the statistical properties of variety $\sigma(t)$. In the bottom panel
of Figure 2 we show the variety of a surrogate market generated according to 
Eq. (6). The time series is very different 
from the real one showed in the top panel of the same Figure.

\section{Conclusion}

The present study performs an analysis the dynamics of returns of an 
ensemble of 
stocks traded in the NASDAQ. We observe that also in the NASDAQ 
the variety is a long-range correlated stochastic variable and that 
time-averaged and ensemble-averaged price returns have different 
statistical properties. Therefore the empirical results found in the NYSE
are not specific of that market but are observed also in the NASDAQ.
The statistical properties of the variety could be  
universal features of financial markets. 
In previous papers \cite{variety00,liegi} we showed that the statistical 
properties of the variety cannot be explained by the single-index 
model. 
A theoretical challenge is to find a model able to explain these  
empirical ensemble observations.

\begin{ack}
The authors thank INFM and MURST for financial support. This work 
is part of the FRA-INFM project {\it Volatility in financial markets}. 
F. Lillo acknowledges FSE-INFM for his fellowships.
We wish to thank Giovanni Bonanno for help in numerical calculations.
\end{ack}

\end{document}